# Optically-powered Low Power Low Noise Amplifiers for MRI

Reza Aghabagheri[1*], Jakob Gerlach[1], Zining Liu[1], Morteza Teymoori[2], Çağlar Ataman[2], Michael Bock[1], Ali Caglar Özen[1]

[1]*Division of Medical Physics, Department of Diagnostic and Interventional Radiology, Medical Center – University of Freiburg, Faculty of Medicine – University of Freiburg, Freiburg, Germany*

[2]*Microsystems for Biomedical Imaging Group, Department of Microsystems Engineering, University of Freiburg, Freiburg, Germany*

**2798**/2800 Words

*Correspondence:

Reza Aghabagheri

Division of Medical Physics, Department of Diagnostic and Interventional Radiology, Medical Center – University of Freiburg, Faculty of Medicine – University of Freiburg, Freiburg, Germany.

Email: Reza.aghabagheri@uniklinik-freiburg.de

# ABSTRACT

**Purpose:** Fully optical receive coils can potentially allow dense receiver arrays with a large channel count, reduced channel crosstalk, and less cable clutter. The power requirements of conventional low-noise amplifiers (LNAs) are prohibitive for simultaneously driving many coils through optical means, as opto-electric power conversion efficiencies can only reach about 50%. The goal is to develop low-power LNAs (LPLNA) with substantially lower power consumption without compromising noise figure (NF) and gain.

**Methods:** A LPLNA was designed as a two-stage cascaded amplifier using an MR-compatible E-pHEMT (Enhancement-mode Pseudomorphic High Electron Mobility Transistor) transistor. The design was implemented on a single-sided printed circuit board (PCB), and its performance was compared with a commercial LNA. A four-channel shielded loop resonator array was constructed, and the signal-to-noise ratio (SNR), noise covariance, and preamplifier decoupling performance were evaluated.

**Results:** The LPLNA had a five-fold lower electrical power consumption (40 mW) than the commercial LNA and provided comparable SNR in phantom measurements. In vivo experiments further confirmed that the LPLNA operates reliably under realistic MRI conditions. Additionally, four-channel receiver array measurements demonstrated comparable SNR within 2% of the commercial LNA and lower inter-channel noise correlation with 0.26 vs 0.3 on average.

**Conclusion:** This study demonstrates the feasibility of LPLNAs for optically-powered RF receiver coil arrays. The LPLNA could also be applied in power-constrained or remote MRI environments.

Word count: 233

## INTRODUCTION

In conventional MRI receive coils the induced signal is inherently weak and must be amplified before it is transferred to the subsequent receiver electronics. For this reason, a preamplifier circuit with a low noise figure (NF) is placed close to each coil element [1]. This low noise amplifier (LNA) increases the signal level, but it should add only minimal additional noise to preserve the signal to noise ratio (SNR). Various LNA designs have been introduced in research settings for MRI applications, with reported gain values ranging from 18 to 32 dB and NF down to sub-1 dB levels, often achieved using E-pHEMT(Enhancement-mode Pseudomorphic High Electron Mobility Transistor)-based architectures and, in some cases, cryogenic operation to further enhance SNR [2,3], [4]. In multi-channel MRI receive arrays, low- or high-input-impedance LNAs can also be used to decouple the coil elements [5–7], which is essential for parallel imaging performance [8]. A range of commercial LNAs are also available for MRI applications including but not limited to [9–12], they typically provide a gain of 20-28 dB with typical NF values around 0.5 dB. Most MRI preamplifiers operate at supply voltages of 5 V or 10 V, with a driving current of 20 to 50 mA. Electrical power to the LNAs is typically supplied through the MRI system interface via a galvanic connection to an external power source. Therefore, the power consumption has not been a design constraint so far.

Recently, we proposed a fully optical signal and power transmission system for MRI coils ("Light Coils") [13–20], to overcome coil coupling, to reduce cable clutter and to provide a modular coil concept with adaptable coil configurations. In these LightCoils, conventional LNAs cannot be used as the power consumption of 200 mW makes wireless [21,22] or optical power delivery impractical. For example, a 64-channel Light Coils array would require approximately 13 W power which is hardly achievable with off-the-shelf optical components. While low-power, high-performance LNAs exist in telecommunications, they generally do not operate optimally at the $^{1}$H Larmor frequency at clinical field strengths, and they are often not MR compatible.

Here, we introduce a low-power, low-noise amplifier (LPLNA) which is designed to work at low bias voltages and powered by a single photonic power converter (PPC). The LPLNA uses a two-stage cascaded design and it was optimized for low power consumption without compromising signal gain and NF. The preamplifier was designed for both 50 Ω input impedance and low input impedance configurations. We also demonstrate the feasibility a four-channel receiver coil array interfaced with LPLNAs.

# METHODS

## LPLNA Design

A dual-stage cascaded LNA design was chosen using two E-pHEMT transistors, where the output of the first stage feeds into the input of the second (Fig. 1A). The SAV-551+ transistor (Mini-Circuits, Brooklyn, NY, USA) was selected as it can operate with a low voltage of about 1 V and a drain current $I_{drain}$ < 20 mA corresponding to a power consumption of $P$ < 20 mW per transistor. In addition, the transistor provides low NF values at lower frequencies: according to the manufacturer's specification, the transistor has a NF < 0.5 dB at 100 MHz at room temperature.

In Figure 1A the LPLNA schematic is shown, components in blue represent the bias network of the circuit. The DC voltage from the PPC (here shown as a voltage source) is supplied to the transistors Q1 and Q2 via choke inductors L4 and L8, respectively. R1 and R2 form a voltage divider that sets the gate voltage ($V_{gate} \approx \left(\frac{R_1}{R_1+R_2} * V_{dc}\right)$) while the capacitor C5 filters out voltage ripples on the gate supply. R1 and R2 are chosen to be sufficiently large to minimize current consumption through the resistors. A similar biasing network is implemented in the second stage (Q2) using R4, R5, L5, C6, and L7. The two stages are biased through separate networks to allow independent control of the power consumption in each stage.

Capacitors C3 and C10 are used to suppress oscillations at the transistor outputs - their values must be large enough to effectively damp oscillations, yet small enough not to degrade the amplifier gain. To assess the stability of the circuit, input and output stability factors $\mu_1$ and $\mu_2$ were calculated from the S-parameters:

$$\mu_1 = \frac{1-|S_{22}|^2}{|S_{11}-S_{22}{}^*(S_{11}S_{22}-S_{12}S_{21})|+|S_{12}S_{21}|} \quad (1)$$

$$\mu_2 = \frac{1-|S_{11}|^2}{|S_{22}-S_{11}{}^*(S_{11}S_{22}-S_{12}S_{21})|+|S_{12}S_{21}|} \quad (2)$$

In general, an LNA remains stable under all source and load conditions if both $\mu_1 > 1$ and $\mu_2 > 1$. Note that the stability factor of a transistor alone is often less than one; therefore, resistors R3 and R6 were introduced. These resistors were added to the matching and stabilization network, and this stabilization comes at the expense of performance. In particular, according to Friis equation for cascaded noise [23], R3 at the output of the first stage has a larger impact on the overall NF. Therefore, its value was optimized to be just sufficient to ensure stability while minimizing its contribution to NF.

The network formed by C1, C2, and L1 matches the impedance of the coil to the input impedance of Q1, making it an important factor for NF performance [4]. The optimum source reflection coefficient ($\Gamma_{\mathrm{opt}}$) at the transistor input is adjusted to achieve the lowest possible NF. As shown in the circuit schematic in Figure 1A, a T-network topology was used. In this implementation, L1 was realized by a high-quality factor air core inductor with a low series resistance to minimize the NF.

All component values are listed in Table 1. Two implementations were realized. In Case A, the input matching network was designed for a 50-Ω source impedance while minimizing the NF, and this configuration was used for all single-channel measurements. For the four-channel measurements and evaluation of preamplifier decoupling, the LPLNA was operated in the Case B configuration, which reduced its input impedance to 5 Ω and enabled comparison with a commercial LNA having an input impedance of 3 Ω. The LPLNA was simulated (ADS 2025, Keysight Technologies, Santa Rosa, CA, USA) to estimate gain, NF, stability, power consumption, and input and output impedances.

## Bench Measurements

For circuit construction a single-sided printed circuit board (PCB) was fabricated using 35-μm-thick copper layer with tin-plating. S-parameter were measured using a network analyzer (P5020A, Keysight Technologies, Santa Rosa, CA, USA) over a frequency range of 110–170 MHz and compared with those of a vendor-supplied commercial LNA. The stability factors $\mu_1$ and $\mu_2$, and the 1 dB gain compression point, $P_{1dB}$, were determined. NF and gain measurements were performed using a signal analyzer (N9000B, Keysight Technologies, Santa Rosa, CA, USA) with a series noise source (N4000A, Keysight Technologies, Santa Rosa, CA, USA). To reduce the effect of external noise, the preamplifier circuit was placed in a shielded box.

Prior to the four-channel MRI measurements, preamplifier decoupling was evaluated to assess the isolation between adjacent coils achieved by the LNA design. For this purpose, two identical coils were placed and fixed next to each other on top of a phantom, and the transmission coefficient (S21) was measured under three conditions: with the coils directly connected to a network analyzer, connected to the input of the LPLNAs, and connected to commercial LNAs.

## Optical Powering

To optically power the LPLNA, an 850 nm laser source (RTMDL-852-1W, Roithner Lasertechnik, Vienna, Austria) was used, which was placed outside of the MRI room. The light was transferred via a multimode optical fiber (MM 50/125 OM2, FS, New Castle, DE, USA) to the LPLNA circuit, where a single-junction GaAs-based PPC converted the incident optical power into electrical power[24]. To assess the optical power requirements, the laser output power was varied between 10-160 mW, and the gain and the NF of the LPLNA were assessed in comparison with a DC power supplied setting (1.2V and 33 mA).

## MRI Measurements

To compare the performance of the LPLNA with a commercial LNA, a shielded loop resonator (SLR) [25–36] with an active detuning network was constructed and mounted on a vendor-supplied bottle phantom ($T_1$ = 132±21 ms, $T_2$ = 47±10 ms). The SLR had a resonance frequency of 123.2 ± 0.2 MHz and an unloaded Q-factor of 70±10. For direct pixel-by-pixel comparison, either the commercial LNA or the LPLNA was connected to the SLR. Phantom measurements were acquired at a clinical 3T MRI system (MAGNETOM CimaX, Siemens Healthineers) using a 2D FLASH sequence (TR/TE = 25/10ms, FA = 10°, ΔV = 1.17 × 1.17 × 5 $mm^3$). To find the optical power setting where commercial LNA and LPLNA yielded similar results, optical power levels between 60 and 150 mW were used to supply the LPLNA.

In addition, to evaluate the compatibility with realistic loading and imaging conditions, in vivo MRI measurements were performed at 3 T using an optically powered LNA operated with 100 mW optical input power. Three imaging protocols were acquired in sagittal orientation: a 3D T1-weighted MPRAGE sequence (TR/TE = 2300/3.04 ms, FA = 8°, ΔV = 0.86 × 0.86 × 0.90 $mm^3$), a 2D T2-weighted turbo spin-echo (TSE) sequence (TR/TE = 4000/79 ms, FA = 150°, ΔV = 0.86 × 0.86 × 3.00 $mm^3$), and a 2D gradient-echo sequence (TR/TE = 186/4.21 ms, FA = 50°, ΔV = 0.86 × 0.86 × 4.00 $mm^3$).

Phantom and in vivo MRI measurements were also performed on a four-channel coil array (Supporting Information Fig. S1) to evaluate the repeatability of the LPLNA and its performance in a multi-channel configuration. In phantom measurements, SNR distribution and noise correlation was benchmarked against the commercial LNAs. A battery-supplied power hub (Supporting Information Fig. S2B) was constructed to generate equivalent voltage and current ratings of 1.2 V and 33 mA to all four channels. Subsequently, four-channel in vivo MRI measurements were performed using the same GRE and T2-weighted TSE protocols as described for the single-channel in vivo measurements, with additional two-fold accelerated acquisitions using GRAPPA [37]. During the in vivo measurements, the LPLNAs were powered optically using 700 mW total laser power distributed to two PPCs, each driving two LPLNAs.

## RESULTS

In the simulation the LPLNA achieved a gain = 31 dB, NF = 0.7 dB, $Z_{in}$ = 5 Ω, Zout = 50 Ω, and unconditional stability while consuming only 30 mW of power (Fig. 1). The simulation shows that the selected transistor can operate at a low bias point of 1 V and 13 mA (Supporting Information Fig. S4A), which is compatible with the maximum output voltage of the PPC, 1.1 V (Supporting Information Fig. S4B).

At bench measurements, with an electrical voltage supply of 1.2 V at 33 mA, LPLNA had a NF = 0.6 dB and gain = 30 dB. The PPC supplied with 100 mW optical power could deliver 1.1 V at 35 mA, leading to a NF = 0.6 dB and gain = 27 dB (Fig. 2A, B). Figure 2C shows that the P1dB was −25 dBm and -12 dBm for the LPLNA and commercial LNA, respectively. The commercial LNA had a NF = 0.7 dB and gain = 25 dB with 10V voltage supply at 20 mA. Stability factors calculated from the measured S parameters indicated unconditional stability, with the LPLNA exhibiting a higher stability margin than the commercial LNA (Fig. 2E, F).

Figures 3A and 3B show single-channel coil SNR maps of the LPLNA (powered by the power hub at 1.2 V and 33 mA) and the commercial LNA, respectively. In the SNR ratio map (Fig. 3E) the LPLNA achieved 4% higher SNR at 40 mW electrical power, fivefold lower power than the commercial LNA. Figures 3F–3H show results for optical powering: at 100 mW of optical input power, the LPLNA achieved an SNR comparable to that of the commercial LNA, and at 60 mW optical power, the SNR was only 4% lower than the commercial LNA. Increasing the optical input power to 150 mW results in an SNR was 3% higher than the commercial LNA.

Figures 3I-3K present the in vivo MR images acquired in a volunteer using the optically powered LPLNA. The T1-weighted MPRAGE, T2-weighted TSE, and FLASH imaging protocols showed the expected contrast in the occipital brain region.

For the four-channel measurements, the input matching network was modified (Case B configuration in Table 1), resulting in a NF of 0.9 dB and a slight decrease in gain to 26 dB at the same power level (40mW electrical power). The additional decoupling through the LPLNA between two adjacent coils was -12 dB, which was 2 dB higher than with the commercial LNA (Fig. 4A). Average noise correlation for the LPLNA was 0.26, and 0.30 for the commercial LNA, indicating a slightly improved isolation performance of the LPLNA (Fig. 4B). Figures 3C and 3D shows four-channel SNR maps. The SNR profiles are shown in Figures 3E and 3F. Across profiles, the mean SNR difference between the LPLNA and commercial LNA was less than 2%.

Figures 4G-4I present four-channel in vivo MR images acquired using the optically powered LPLNA-based coil array. The successful visualization of anatomical structures across the four-channel acquisition demonstrates that the proposed LPLNA design can be extended from a single-channel implementation to a multi-channel receive configuration while maintaining optically powered operation.

A long-term stability measurement was conducted (Supporting Information Fig. S5) by repeating the experiment 512 times over approximately 30 minutes. The SNR, mean signal, and noise standard deviation were calculated within a selected ROI and plotted together. The mean SNR of the LPLNA was 11% higher than that of the commercial LNA, with only 4% average variation from the corresponding mean SNR.

# DISCUSSION

In this work, a two-stage LPLNA was designed and tested to enable optical powering in a multi-channel coil setup. The LPLNA consumed 40 mW of electrical power, so that it could be driven by a single PPC. Despite this significant reduction in power, performance was comparable to the commercial LNAs and it was compatible with various in vivo pulse sequences

A limitation of the LNLPA design is that the E-pHEMT transistors are sensitive to their orientation to the magnetic field: at the center of the magnet bore, the circuit needed to be rotated by 90° to the magnetic field for optimum performance – a feature that was also observed in some commercial LNAs. In general, this is not a strong limitation if the circuit can be arranged in a fixed way relative to the coil, as the coil needs to have a certain orientation for highest sensitivity.

With the chosen off-the-shelf transistor, the power consumption could be reduced to 40 mW so that a 64-channel coil could be operated with only 2.6 W of electrical power. Assuming an optical-to-electrical power conversion efficiency of 50% as in this study, a 64-channel coil array would require only 5.2 W of optical power. However, as observed in the optically-powered four-channel in vivo measurements, interconnections introduced in the fiber optic chain resulted in additional losses and total optical power was 700 mW instead of 400 mW. Any extrapolation to large arrays, therefore, must also consider system level constraints. Power levels up to 10 W can realistically be supplied using commercially available components. Further increasing the number of channels would either require additional power optimization of the LPLNA (e.g., integrated circuit implementation) or the use of additional fibers and laser sources.

The input matching network of the LPLNA can be modified to support different coil configurations and operating frequencies, including those used for X-nuclear imaging. In addition, the stable gain response above 100 MHz indicates potential applicability at higher field strengths.

In our experiments, the laser source powering the LPLNA was not gated or triggered by the MR sequence but instead operated continuously. This continuous-power strategy avoids delays associated with the rise time of the electronics when switching back to receive and helps prevent transient signal loss at the beginning of signal reception. In our MRI interfaces, commercial preamplifiers also remain active during the transmit phase, including those connected to receive coil elements that are not selected for signal reception.

## CONCLUSION

A LPLNA for MRI with optical powering was presented that achieves comparable performance to a commercial LNAs while reducing the electrical power consumption by five-fold. The LPLNA uses a two-stage cascaded design based on an MR-compatible transistor and an optimized matching and stabilization network, offering low NF, sufficient gain, and unconditional stability under realistic operating conditions. Bench measurements confirmed the simulated performance in terms of gain, NF, stability, and 1 dB compression, and demonstrated that the amplifier can be efficiently powered by a single PPC using an 850 nm laser source. The LPLNA was studied in single-channel phantom and in vivo MRI experiments, and in a four-channel coil array. Results indicate that this LPLNA design could be used for fully optically driven receive arrays. Further power optimization and potential integrated-circuit implementations could enable even higher channel counts and expand the applicability of optically powered MRI technology to mobile and resource-constrained environments.

## ACKNOWLEDGMENTS

This study was funded by Deutsche Forschungsgemeinschaft (DFG, German Research Foundation) under grant number #1050092401.

# FIGURES

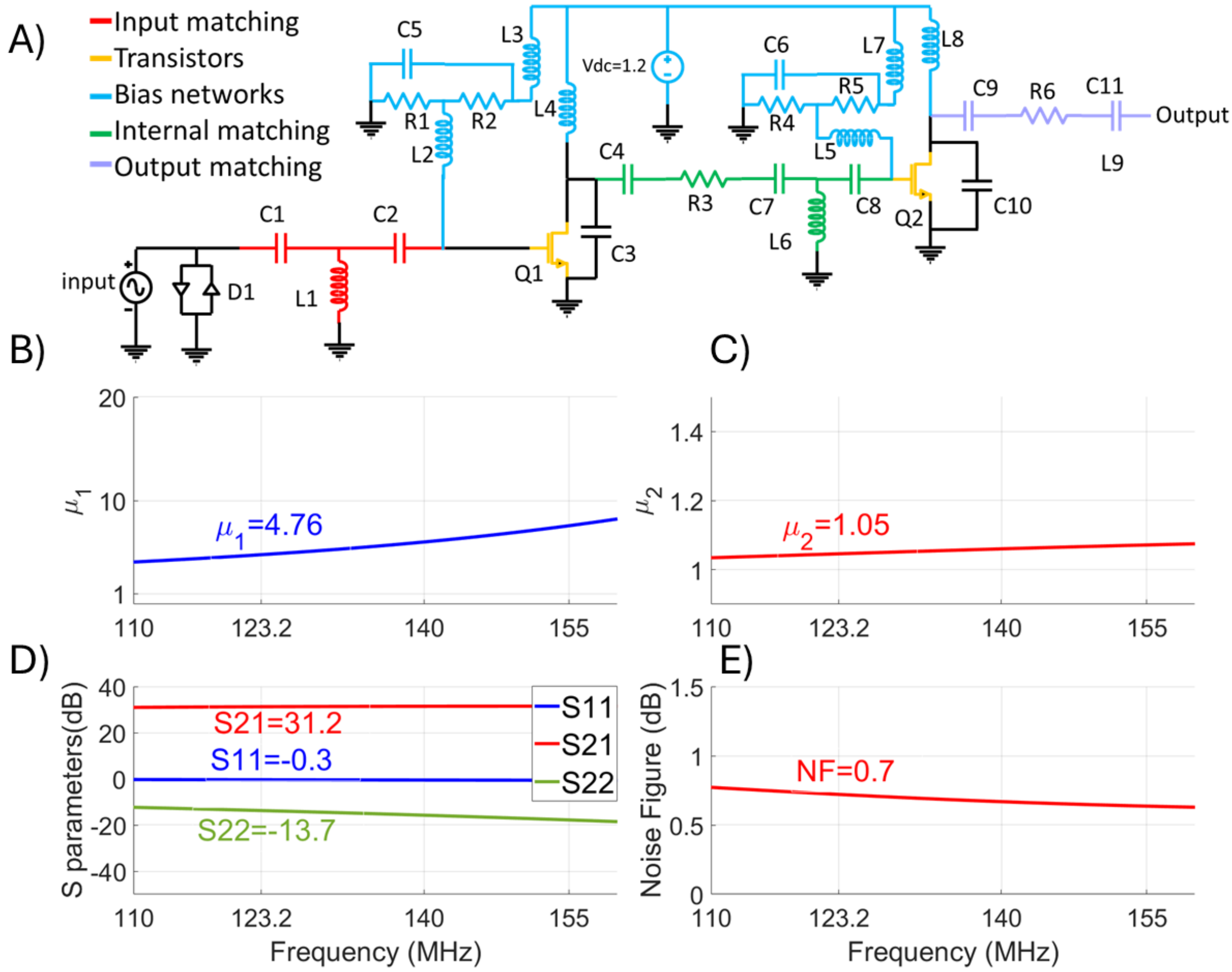

Fig. 1:(A) Schematic of the LPLNA circuit design. RF circuit simulations show unconditional stability with stability factors $\mu_1$ (B) and $\mu_2$ (C) larger than 1. (D) S-parameters show gain, low input impedance and sufficient output matching. (E) Noise figure was also simulated and was below 0.7 dB for the given input matching.

Table 1: List of components with their respective values.

| **Components** | **Values** | |
|---|---|---|
| **D1** | Cross Diode | |
| **C1** | 5-30 pF | |
| **L1** | Case A | Case B |
| | 100 nH | 47 nH |
| **C2** | Case A | Case B |
| | 20 pF | 1 nF |
| **Q1, Q2** | SAV-551+ | |
| **C3, C10** | 39 pF | |
| **C4,C5,C6, C8, C9, C11** | 10 nF | |
| **R3, R6** | 56 Ω | |
| **C7** | 10 pF | |
| **L6** | 110 nH | |
| **R1, R4** | 330 kΩ | |
| **R2, R5** | 680 kΩ | |
| **L2, L3, L4, L5, L7, L8** | 8.2 uH | |

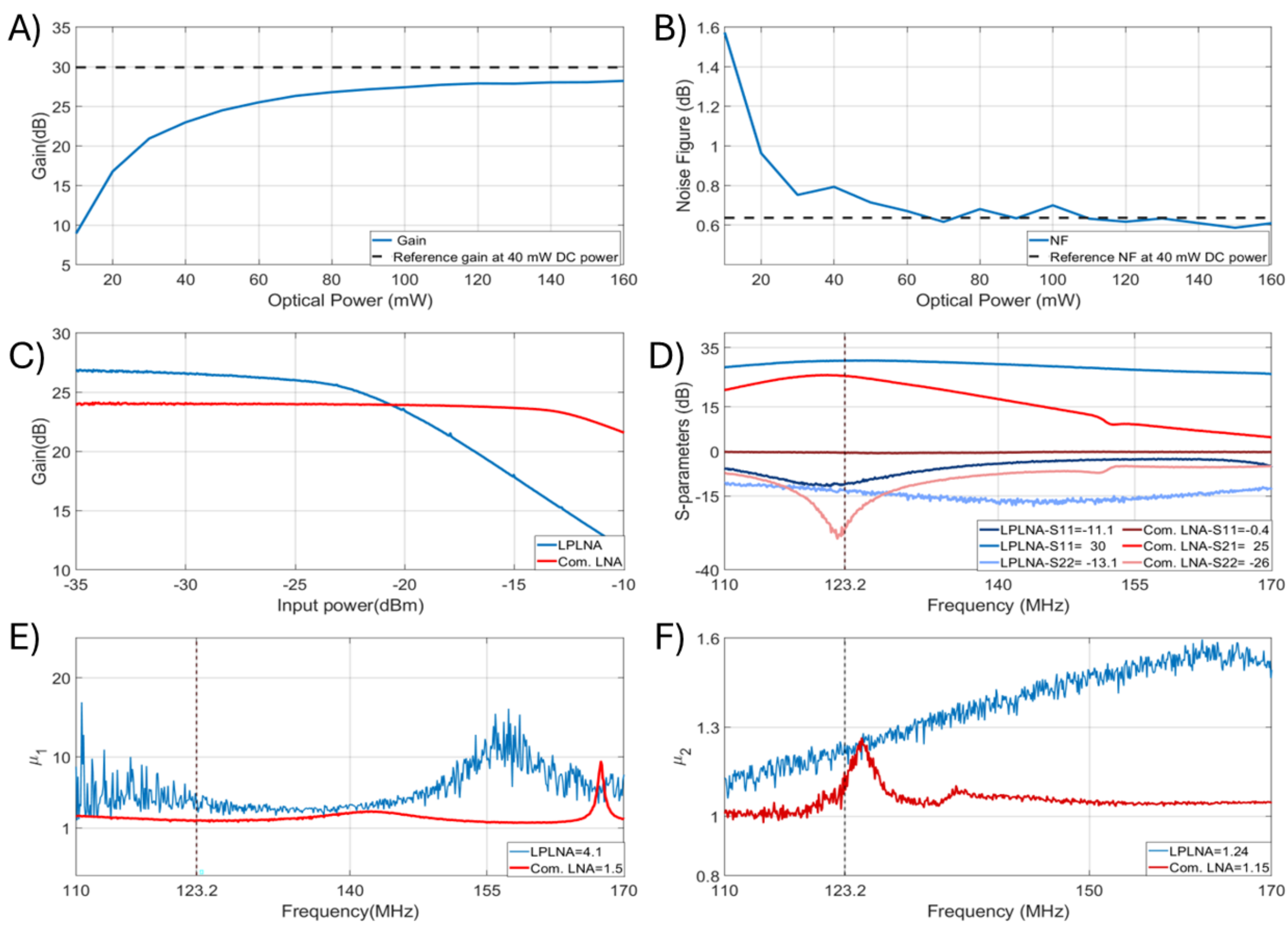

Fig. 2: Bench tests for optically powered LPLNA. (A) Gain and (B) NF vs. optical power at the PPC. The reference line (black, dashed) represents the gain and NF at 40 mW DC power (1.2 V and 33 mA). At 100 mW of optical power 27 dB of gain and 0.6 dB of NF are reached. (C) Input RF power vs. gain for LPLNA and commercial LNA. 1 dB suppression points are at -25 dBm and -12 dBm for LPLNA and commercial LNA, respectively. (D) S-parameters of LPLNA and commercial LNA. (E) $\mu_1$ and (F) $\mu_2$ stability factors calculated from S-parameter measurements.

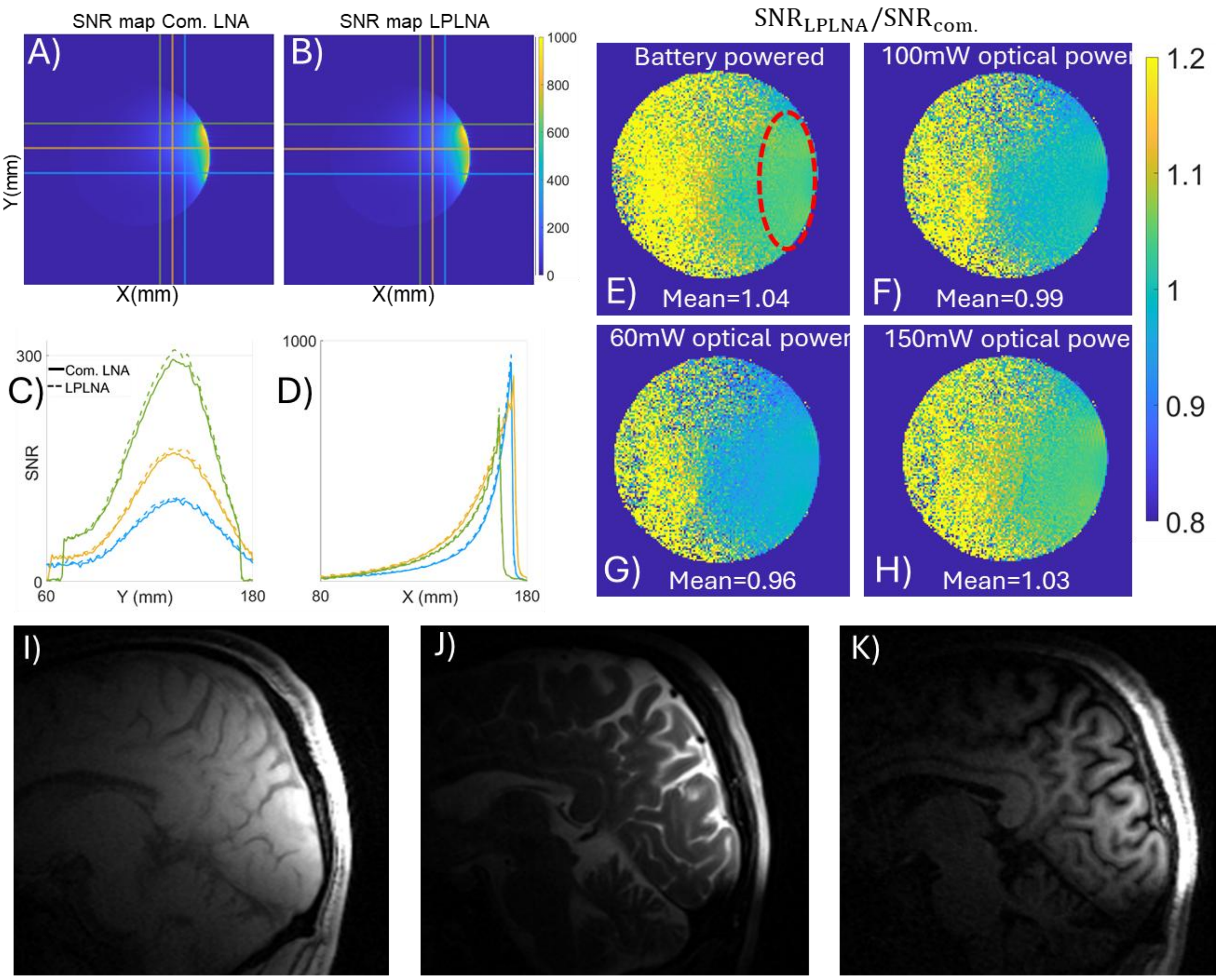

Fig. 3: (A) SNR map acquired with a commercial LNA with selected rows and columns highlighted for line profile analysis. (B) Corresponding SNR map with the LPLNA. (C) Column SNR profiles for the selected columns (D) Row SNR profiles for the selected rows using the same color coding. SNR map ratio (E) in a case that 40mW electrical power comes through power hub. (F, G, H)) Optical power fed through PPC to the LPLNA. Single-channel in vivo sagittal images: (I) T1w spoiled 2D FLASH (J) T2-weighted TSE, and (K) T1-weighted MPRAGE.

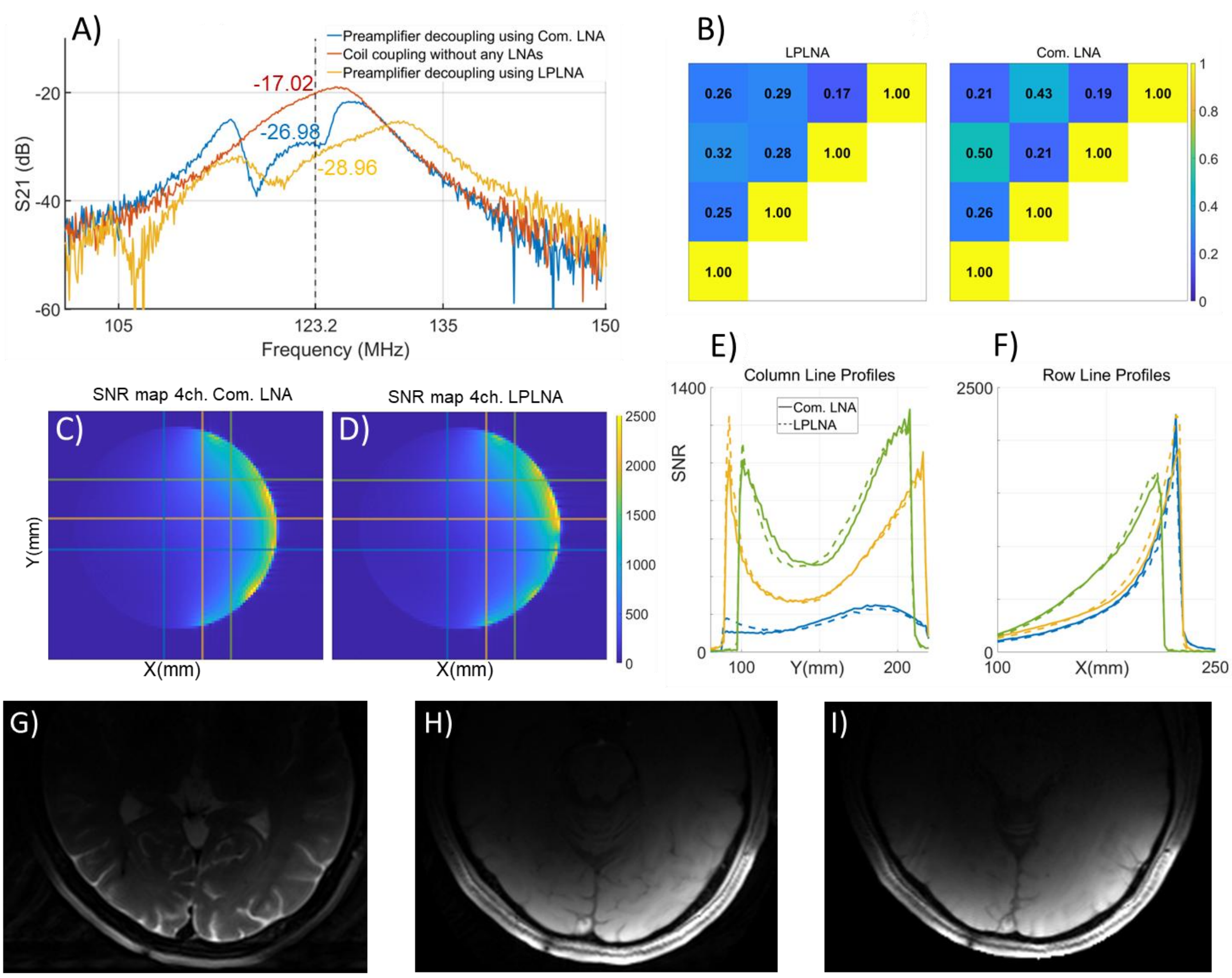

Fig. 4: (A) Preamplifier decoupling bench measurement result. (B) Noise correlation matrices. (C) Four-channel SNR map with commercial LNAs and selected rows and columns. (D) Corresponding four-channel SNR map with the LPLNA. (E) Column and (F) Row SNR profiles for the four-channel SNR map. 4-channel in vivo transverse images: (G) T2-weighted TSE, and 2D FLASH without (H) and with 2-fold GRAPPA acceleration (I).

# Supplementary INFORMATION

# Optically-powered Low Power Low Noise Amplifiers for MRI

Reza Aghabagheri[1*], Jakob Gerlach[1], Zining Liu[1], Morteza Teymoori[2], Çağlar Ataman[2], Michael Bock[1], Ali Caglar Özen[1]

*[1]Division of Medical Physics, Department of Diagnostic and Interventional Radiology, Medical Center – University of Freiburg, Faculty of Medicine – University of Freiburg, Freiburg, Germany*

*[2]Microsystems for Biomedical Imaging Group, Department of Microsystems Engineering, University of Freiburg, Freiburg, Germany*

*Correspondence:

Reza Aghabagheri

Division of Medical Physics, Department of Diagnostic and Interventional Radiology, Medical Center – University of Freiburg, Faculty of Medicine – University of Freiburg, Freiburg, Germany.

Email: Reza.aghabagheri@uniklinik-freiburg.de

# SUPPORTING INFORMATION

## Four channel coils

As described above, the performance of the LPLNA design was evaluated in a four-channel receive coil array. For this purpose, four shielded loop resonators were taped on a leather substrate. As shown in Supporting Information Figure S1, geometric overlap was used to decouple the individual coil elements. The measured inter-element coupling ranged from −27 dB to −11 dB across the array, while the reflection coefficients of the individual channels, S11–S44, were maintained at −15 ± 3 dB.

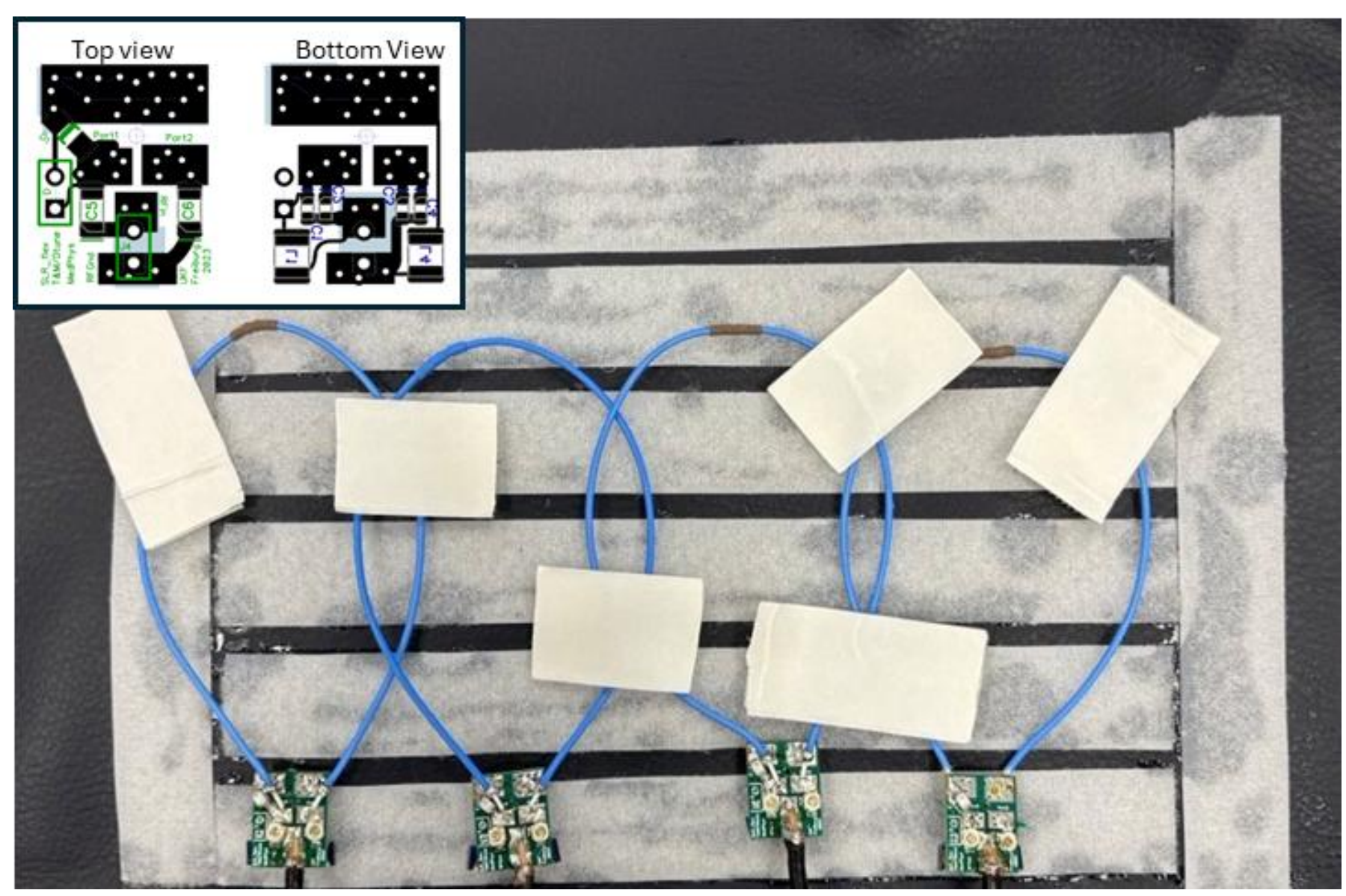

Supporting Information Figure S1: Photograph of the fabricated 4-channel coil array geometry decoupled used for the measurements, together with the PCB layout overview of the tuning and matching circuit.

## PCB Layouts

As shown in Supporting Information Figure S2A, the LPLNA was implemented on a 43 mm × 44 mm single-sided PCB. All electronic components were mounted and soldered on the top side of the PCB.

The power hub generates multiple output voltage levels from a single input source, enabling the evaluation of different LNA supply voltages inside the MRI room using one battery. It also allows several LNAs to be powered simultaneously. The ADPL44002 (Analog Devices, Wilmington, MA, USA) family which is low-dropout linear commercial regulator is used for this purpose.

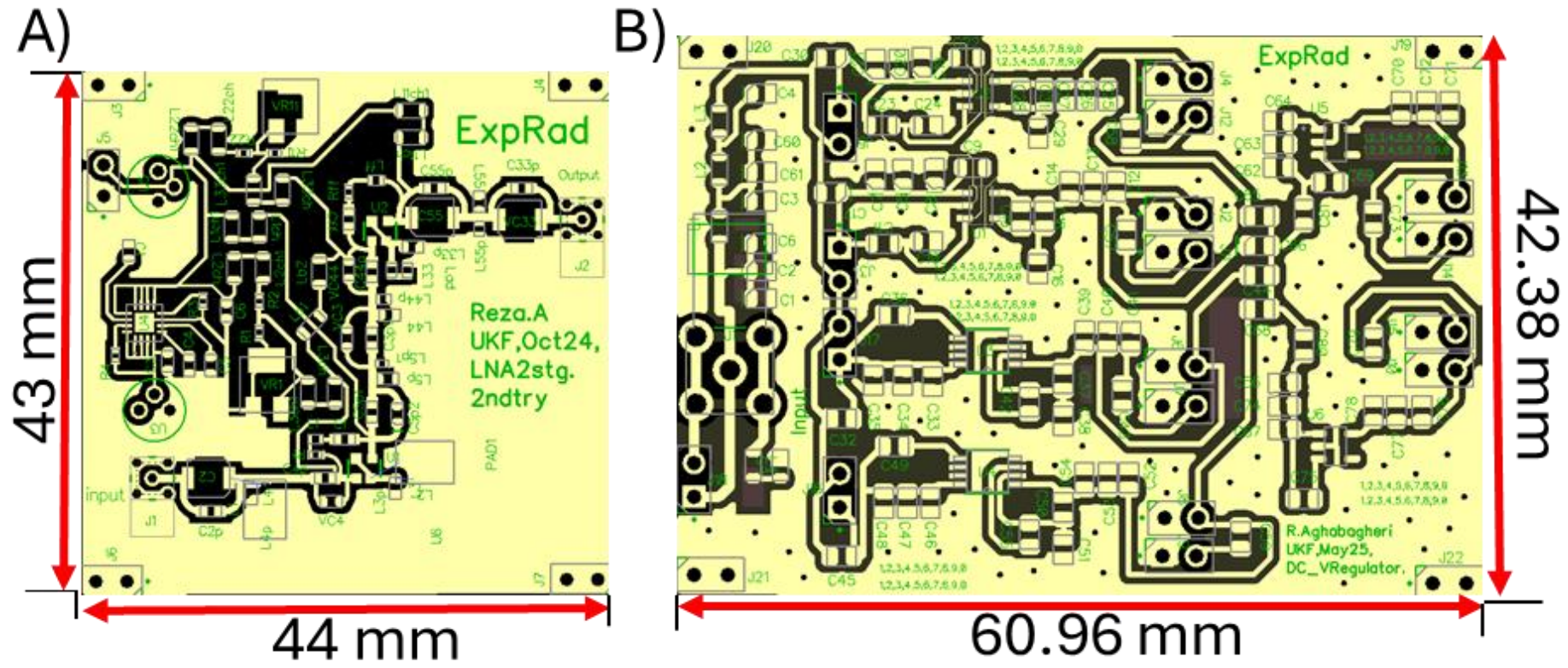

Supporting Information Figure S2: (A) An overview of the PCB layout designed for the LPLNA. (B) PCB layout of the power hub designed to supply multiple LNAs.

A T-circuit was designed and used to deliver the detuning signal directly to the coil while suppressing the RF signal in the DC path. At the same time, the signal from the coil is routed to the preamplifier with low insertion loss. As shown in Supporting Information Figure S3B, the DC path at 123.2 MHz provides more than 30 dB isolation, whereas the signal path has only 0.1 dB loss.

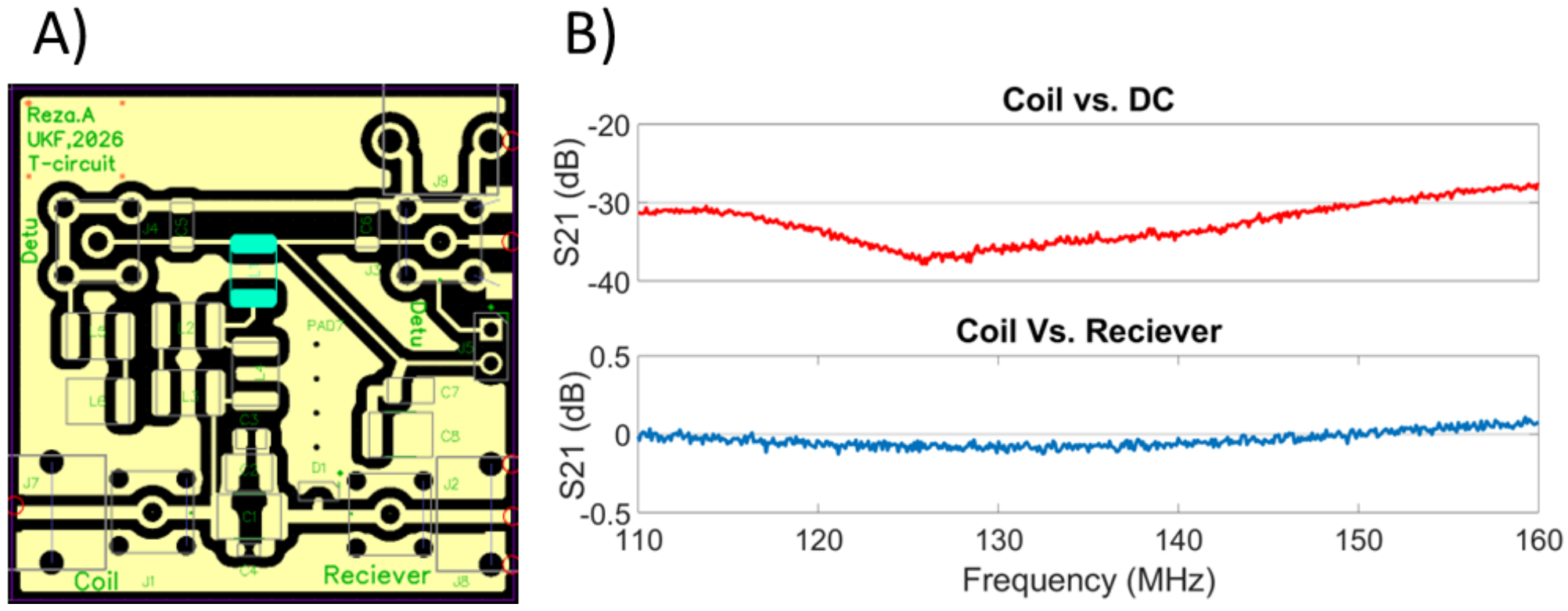

Supporting Information Figure S3: (A) An overview of the PCB layout designed as a T-circuit. (B) S21 measurement evaluating isolation between coil and DC path and loss between coil and the receiver path.

## Power Budget and Bias Requirements

The minimum electrical power required for the selected transistor to operate in amplification mode, as well as the maximum electrical power that could be delivered by the photonic power converter (PPC) were evaluated. Supporting Information Figure S4A shows the simulated I-V characteristics of the single transistor. As an example, the transistor can operate as an amplifier at a drain voltage of 1 V and a drain current of 13 mA, which requires a gate voltage of approximately 0.4 V. Supporting Information Figure S4B shows the I-V characteristics of the PPC under different incident optical power levels. For example, at an incident optical power of 100 mW, the PPC can provide a maximum voltage of slightly below 1.2 V and a current of 35 mA. Based on a simple calculation of the generated electrical power, the optical-to-electrical conversion efficiency of the PPC is therefore below 50%.

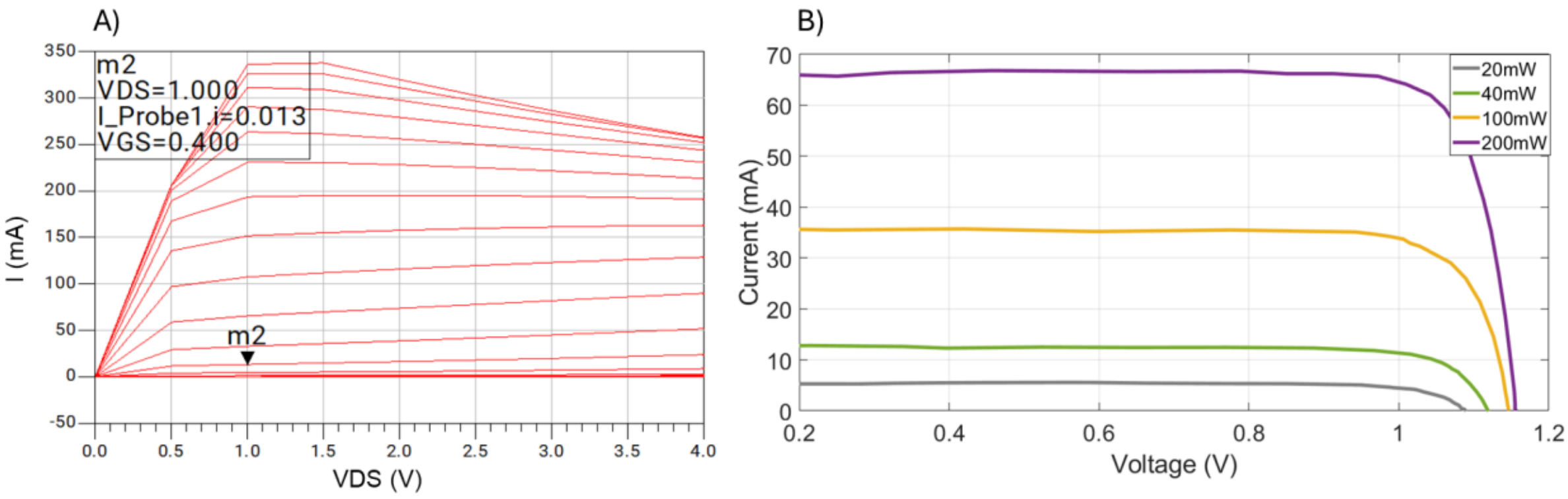

Supporting Information Figure S4: (A) Simulated I–V characterization of the transistor used in the LPLNA design. (B) I-V characterization of the photonic power converter

## Stability Measurements

To evaluate the performance of the LPLNA over time during MRI measurements, a 2D FLASH sequence was repeated continuously 512 times over approximately 30 minutes. The same measurement was subsequently performed using the commercial LNA. For each dataset, the mean signal, noise standard deviation, and mean SNR were calculated over time from selected signal and noise ROIs and plotted together to compare stability and performance between the two amplifiers.

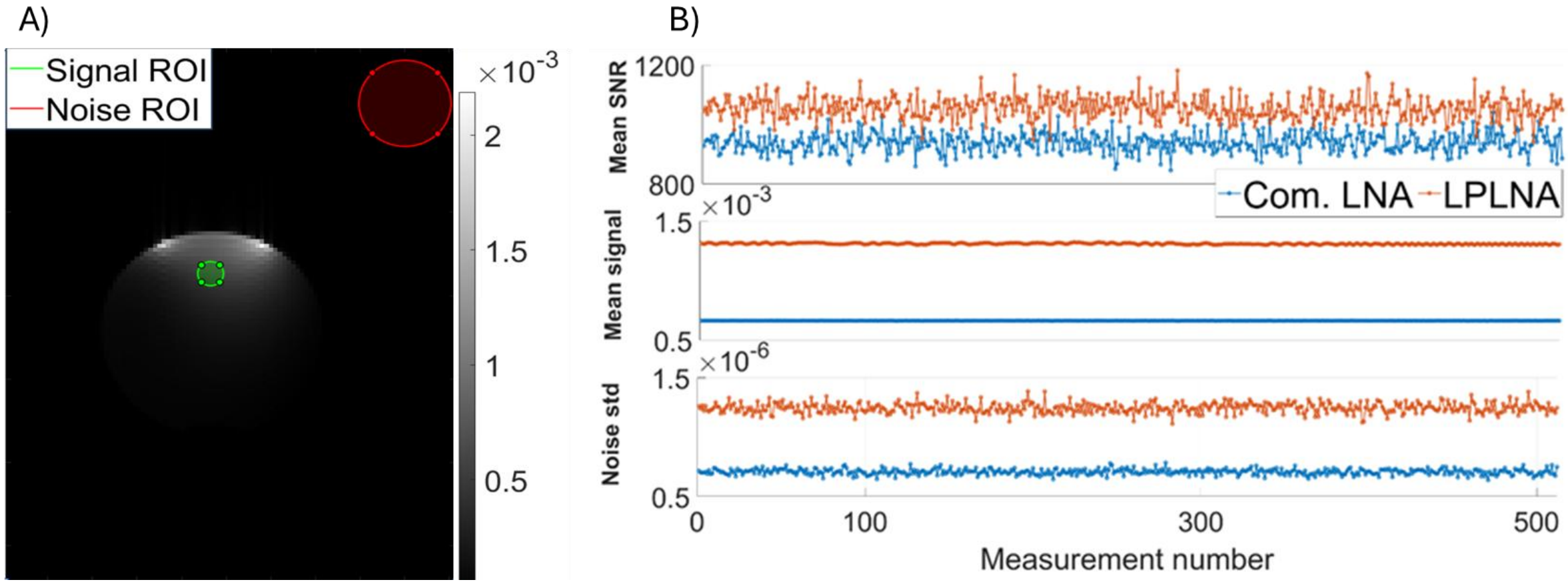

Supporting Information Figure S5: (A) Representative MR image showing the signal ROI placed within the phantom and the noise ROI placed in the background used for SNR calculation. (B) Long-term stability measurement showing the SNR, mean signal, and noise standard deviation